\documentstyle[aps,epsfig]{revtex}
\begin{document}
\draft
\title{Magnetic Anisotropy in the Molecular Complex V$_{15}$}

\author{N.P. Konstantinidis}

\address{Department of Physics and Department of Mathematics, Trinity College,
Dublin 2, Ireland} 

\author {D. Coffey$^{*}$}

\address{Department of Physics, State University of New York at Buffalo,
Amherst, NY 14260}

\date{\today}
\maketitle

\begin{abstract}
We apply degenerate perturbation theory to investigate the effects of magnetic
anisotropy in the magnetic molecule V$_{15}$. Magnetic anisotropy is introduced
via Dzyaloshinskii-Moriya (DM) interaction in the full Hilbert space of the
system. Our model provides an explanation for the rounding of transitions in
the magnetization as a function of applied field at low temperature, from which
an estimate for the DM interaction is found. We find that the calculated energy
differences of the lowest energy states are consistent with the available data.
Our model also offers a novel explanation for the hysteretic nature of the
time-dependent magnetization data. 
\end{abstract}

\pacs{PACS numbers: 75.10.Jm Quantized Spin Models, 75.50.Ee
      Antiferromagnetics, 75.50.Xx Molecular Magnets}

\section{Introduction}

Magnetic molecules have attracted significant attention in recent years as
systems where macroscopic quantum phenomena are displayed, such as relaxation
by tunneling of magnetization through a potential well
\cite{Caneschi}-\cite{Gunther}. These molecules form crystals with very weak
intermolecular interactions, thus experiments reflect directly the properties
of individual molecules, which can be described by magnetic Hamiltonians. The
most studied molecules are a cluster of manganese ions known as Mn$_{12}$
\cite{Caneschi2}, and clusters with iron ions, Fe$_{n}$ \cite{Delfs,San,Cac}.

Mn$_{12}$ and Fe$_{8}$ possess strong easy axis anisotropy and their states can
be characterized to a very good approximation by $<S^{z}>$, the expectation
value of the total spin along the easy axis, $z$. Their ground state has a
total spin $S=10$, and to a good approximation they are described by a
single-spin model, where only the $S=10$ multiplet states are considered
\cite{Leuenberger2,Barra3}. Restriction to the $S=10$ multiplet reduces
drastically the Hilbert space of the problem. Iron atoms also form ferric wheel
clusters with predominantly antiferromagnetic interactions. In these the total
spin of the ground state is $S=0$\cite{Fe6}-\cite{Fe18}. Spin dynamics of
Fe$_{6}$ and Fe$_{8}$ have recently been calculated by Honecker et al.
\cite{Honecker} using exact diagonalization.

Another molecule of the same class is a cluster with $15$ Vanadium ions, known
as V$_{15}$ \cite{Muller3}, each of which has $S = \frac{1}{2}$. This is
similar to some of the Fe$_n$ systems in that the principal interactions are
antiferromagnetic. This system has also been investigated using effective spin
models in which sets of spins are replaced by one effective spin. However,
effective three-spin models which contain only isotropic terms
\cite{Chiorescu}-\cite{Raghu} can not explain the broadening of the spin
transitions as a function of magnetic field in equilibrium magnetization
measurements, as was pointed out by Chiorescu et al. \cite{Chiorescu}.
Miyashita and Nagaosa \cite{Miyashita01} considered a three-spin model for
V$_{15}$ which included a term, $\sum_{ij}\alpha_{ij}S^z_iS^x_j$. This term is
responsible for the quantum tunneling of magnetization states and for the
smoothing out of transitions between magnetization states in applied magnetic
fields. However, a more microscopic model is needed if a quantitative
comparison with data is to be made. In particular, the tunnel splittings, which
are important for magnetic relaxation, are very sensitive to Hilbert space
truncation and the full Hilbert space of the molecule has to be taken into
account \cite{DeRaedt00}. Here we solve for the lowest energy states using
degenerate perturbation theory\cite{Gelfand,NPK} in the full Hilbert space,
which has $2^{15}=32,768$ states.

Apart from the opportunity to study the magnetic properties of finite systems
which these systems provide, Leuenberger and Loss \cite{Leuenberger} have
proposed the use of Mn$_{12}$ as a qubit, a quantum bit, in a quantum computer.
In this type of application it is important that prepared spin states do not
decohere on the time scale of the operations on the system necessary to read,
write and carry out the manipulations necessary for a computation. Consequently
it is important to understand the mechanisms for magnetic relaxation in these
molecules. These mechanisms can be probed by the application of time dependent
magnetic fields \cite{Barbara98,Zhong00}.

The magnetic hysteresis studies for V$_{15}$ show so-called ``butterfly''
curves which are different from the ones of Mn$_{12}$ and Fe$_{8}$
\cite{Chiorescu,Chiorescu2}. This has been explained in terms of spin-phonon
interaction using the Landau-Zener model for transitions between two states
\cite{Stevens}. However, the low number of phonons at very low temperatures
points to a different mechanism for magnetization relaxation, intrinsic to the
molecule. Here we examine the role of magnetic anisotropy described by the
Dzyaloshinskii-Moriya (DM) interactions as a relaxation mechanism. This
interaction arises from spin-orbit coupling and was introduced by
Dzyaloshinskii phenomenologically by using symmetry considerations
\cite{Dzyaloshinskii}. Moriya derived it microscopically by extending the
Anderson theory of superexchange to include spin-orbit coupling \cite{Moriya}.
The DM term arises in low symmetry crystals. The fact that finite systems have
surfaces where symmetry is reduced ensures that the DM term is an important
part of any microscopic description of magnetic molecules \cite{Crepieux}.
There are other anisotropy sources such as dipolar and hyperfine fields which
have very small magnitudes, about $1$ mT and $40$ mT respectively
\cite{Chiorescu}. Magnetic anisotropy has also been modeled in Ising
Hamiltonians by $S^{+}$ and $S^{-}$ terms to reproduce the experimental data
\cite{Leuenberger2,Barra3}. Single site anisotropy terms do not contribute in
V$_{15}$ since $\sigma_{i}^{2}=1$, $i=x,y,z$ for $S=\frac{1}{2}$. Therefore the
DM interaction is the leading order anisotropy term for this system.


The model Hamiltonian, which includes a Heisenberg exchange interaction and a
DM anisotropy term, is solved here with degenerate perturbation theory
\cite{Gelfand,NPK}. It is found that inclusion of the DM terms quantitatively
explains the widening of the spin transition in the equilibrium magnetization
curve. The DM terms are determined quantitatively by comparing with
experimental data of Chiorescu et al. \cite{Chiorescu}, and the resulting
splitting of the low energy groups of states is in agreement with experimental
estimates, $\sim 10^{-2}$ K. Furthermore, it is shown with a model calculation
that the non-equilibrium magnetization curve is intrinsically hysteretic due to
the presence of the DM interaction.  This interaction leads to rapid
oscillations wth frequencies given by its magnitude. The form of the hysteresis
curve depends on when during these oscillations the magnetic field is reversed.

The paper is organized as follows. In section \ref{sec:1} we discuss the
magnetic description of the V$_{15}$ molecule and in section \ref{sec:2} we
introduce the model including the DM interaction determined by symmetry. In
section \ref{sec:3} we present the properties of the model and compare with
experimental data. Finally in section \ref{sec:4} we give our conclusions.

\section{V$_{15}$}

\label{sec:1}
\begin{figure}
\begin{center}
\epsfig{file=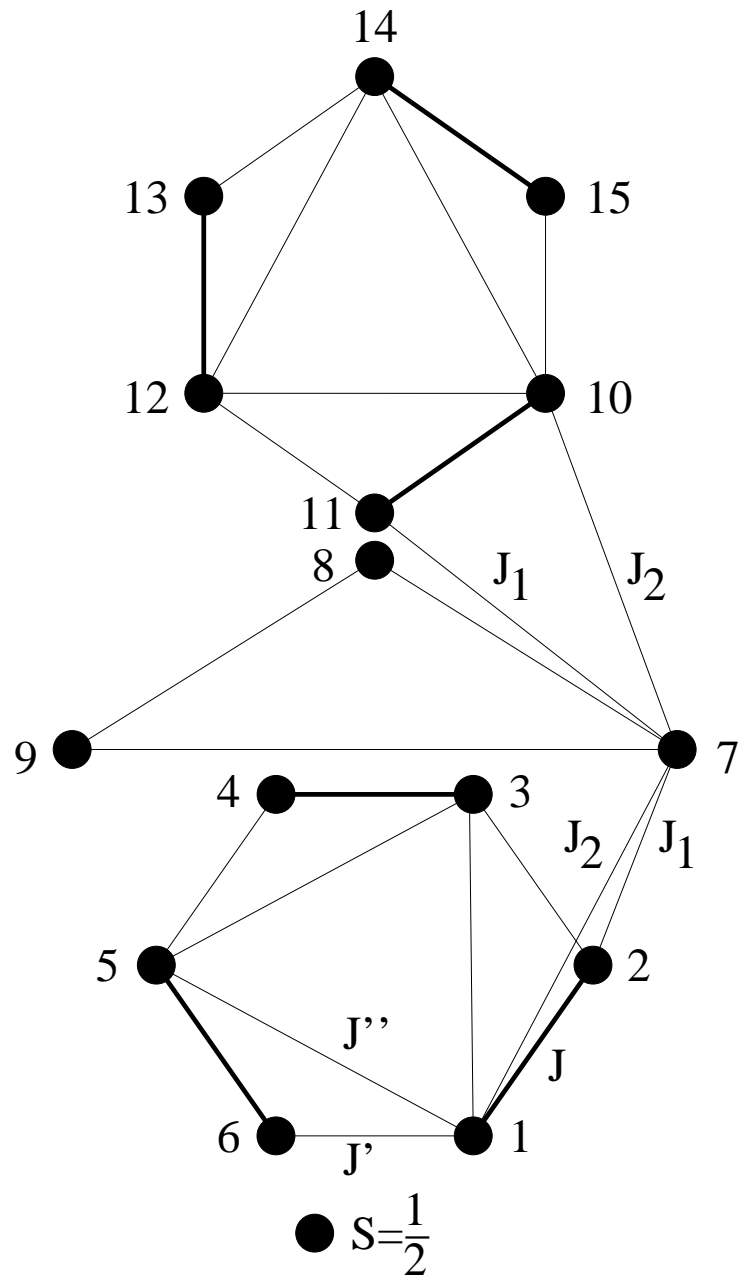,width=3.8in}
\end{center}
\vspace{20pt}
\caption{Space configuration of V$_{15}$.}
\label{fig:1}
\end{figure}

The chemical structure of V$_{15}$ is given by the formula
K$_{6}$ [V$_{15}^{IV}$As$_{6}$O$_{42}$(H$_{2}$O)] $\cdot$ 8H$_{2}$O
\cite{Muller3,Muller,Muller2}. There are fifteen V$^{IV}$ ions with spin
$S=\frac{1}{2}$ placed in a quasi-spherical layered structure (figure
\ref{fig:1}). These spins sit on two hexagons with a triangle sandwiched
between them and all interactions are antiferromagnetic in the simplest model
\cite{Muller3}. The spins in each hexagon are paired via a strong coupling
$J \cong 800$ K \cite{Muller2}. These pairs are connected with a nearest
neighbor interaction $J'\cong 150$ K. There is also a diagonal interaction
$J'' \cong 300$ K. The triangle spins do not interact directly with each other.
They interact with spins from the top and bottom hexagons, and the interactions
are $J_{1} \cong J'$ and $J_{2} \cong J''$. The molecule has trigonal symmetry
and the space group is $R_{\bar{3}c}$. The unit cell includes two molecules but
intermolecular interactions are negligible.

Since all interactions are antiferromagnetic the total spin of the ground state
is $S=\frac{1}{2}$. This is in contrast with Mn$_{12}$ and Fe$_{8}$,
where $S=10$. In the absence of anisotropy the ground state is four-fold
degenerate, and it is made up of two doublets. The lowest lying excited states
form a quartet which is separated from the ground state doublets by
$\approx 3.7$ K. The splitting between the lowest energy multiplets from the
rest of the spectrum has been found from EPR spectra to be $\sim 500$ K
\cite{Barra4}. Magnetic anisotropy leads to tunnel splittings between the two
doublets, which are $\sim 10^{-2}$ K \cite{Carter}, in contrast with the very
small values for the other two molecules ($\sim 10^{-7} - 10^{-11}$ K). There
is no macroscopic quantum tunneling for V$_{15}$ according to the Kramers
theorem \cite{Kramers2}, since the system has an odd number of $S=\frac{1}{2}$
spins and the total spin is half-integer \cite{Bhaduri}, giving a doubly
degenerate ground state.


From equilibrium magnetization measurements, it is clear that the ground state
has $S \cong \frac{1}{2}$ up to a critical value of the field, where a
transition to $S \cong \frac{3}{2}$ is detected \cite{Chiorescu}. This is true
for temperatures smaller than $0.9$ K. The broadening of the spin transitions
can not be attributed to temperature alone, and various factors like dipolar or
nuclear hyperfine field distributions are too small, about $1$ mT and $40$ mT
respectively \cite{Chiorescu}. This broadening can be explained by anisotropic
DM interactions, which were also found to be important for the description of
neutron scattering data and EPR measurements in Mn$_{12}$ \cite{Katsnelson99}.
More recently, analysis of neutron scattering data on V$_{15}$ by Chaboussant
et al. \cite{Chab02} has shown the need to include DM interactions. Garanin and
Chudnovsky \cite{Garanin} have suggested that the width of EPR lines could also
be explained by dislocations in the crystal lattice. Here we assume that there
are no defects in the crystals so that data reflects the properties of the
V$_{15}$ clusters.

\section{Model}

\label{sec:2}
A microscopic treatment of the problem of the origin of magnetic anisotropy is
considered using the full Hilbert space with $2^{15}=32,768$ states. This
approach based on degenerate perturbation theory makes it possible to include
the effect of spin-orbit coupling through the DM interaction. Initially
perturbation theory is applied in the absence of magnetic anisotropy. The
unperturbed Hamiltonian, $H_{0}$, takes into account the spin singlet-triplet
correlations present due to the strength of the bond between alternating pairs
of spins in the hexagons being more than twice as large as any other coupling. 
The ground state of $H_{0}$ is eight-fold degenerate due to uncoupled spins in
the triangle (figure \ref{fig:1}). Then the perturbation includes the remaining
terms for bonds on which exchange constants are smaller. An effective
Hamiltonian, $H^{eff}$, is generated for the degenerate subspace. Since
singlet-triplet correlations of the prevalent $J$ bond are built into $H_{0}$,
the series expansions generated by perturbation theory are absolutely
convergent. Diagonalizing $H^{eff}$ gives the eight lowest energy states of the
system which are separated from the next set of states roughly by the energy
splitting between singlet and triplet states in $H_{0}$. The Hamiltonian is of
the form
\begin{equation}
H = H_{0} + \lambda H_{1}
\label{eqn:1}
\end{equation}
where $\lambda$ is the perturbation parameter with
\begin{equation}
H_{0} = J ( \vec{S}_{1} \cdot \vec{S}_{2} + \vec{S}_{3} \cdot \vec{S}_{4} +
\vec{S}_{5} \cdot \vec{S}_{6} + \vec{S}_{10} \cdot \vec{S}_{11} +
\vec{S}_{12} \cdot \vec{S}_{13} + \vec{S}_{14} \cdot \vec{S}_{15} )
\end{equation}
and
\begin{eqnarray}
H_{1} =
& J' ( \vec{S}_{2}  \cdot \vec{S}_{3}  + \vec{S}_{4}  \cdot \vec{S}_{5}  +
       \vec{S}_{1}  \cdot \vec{S}_{6}  + \vec{S}_{11} \cdot \vec{S}_{12} +
       \vec{S}_{13} \cdot \vec{S}_{14} + \vec{S}_{10} \cdot \vec{S}_{15} ) +
\nonumber \\
& J'' ( \vec{S}_{1}  \cdot \vec{S}_{3}  + \vec{S}_{3}  \cdot \vec{S}_{5}  +
        \vec{S}_{1}  \cdot \vec{S}_{5}  + \vec{S}_{10} \cdot \vec{S}_{12} +
        \vec{S}_{12} \cdot \vec{S}_{14} + \vec{S}_{10} \cdot \vec{S}_{14} ) +
\nonumber \\
& J_{1} ( \vec{S}_{2} \cdot \vec{S}_{7} + \vec{S}_{7} \cdot \vec{S}_{11} +
          \vec{S}_{4} \cdot \vec{S}_{8} + \vec{S}_{8} \cdot \vec{S}_{15} +
          \vec{S}_{6} \cdot \vec{S}_{9} + \vec{S}_{9} \cdot \vec{S}_{13} ) +
\nonumber \\
& J_{2} (\vec{S}_{1} \cdot \vec{S}_{7} + \vec{S}_{7} \cdot \vec{S}_{10} +
         \vec{S}_{3} \cdot \vec{S}_{8} + \vec{S}_{8} \cdot \vec{S}_{14} +
         \vec{S}_{5} \cdot \vec{S}_{9} + \vec{S}_{9} \cdot \vec{S}_{12} )
\end{eqnarray}

The ground state of the unperturbed Hamiltonian is composed of a singlet at
each of the six hexagon pairs. When $\lambda \neq 0$ the perturbation mixes in
states where one or more of the pairs are excited and generate the elements of
$H^{eff}$. Since $J$ is a very strong bond, this lowest multiplet of $8$ states
will be well-separated in energy from the rest of the spectrum, and the low
temperature behavior is well described by the eigenfunctions of $H^{eff}$.

\begin{figure}
\begin{center}
\epsfig{file=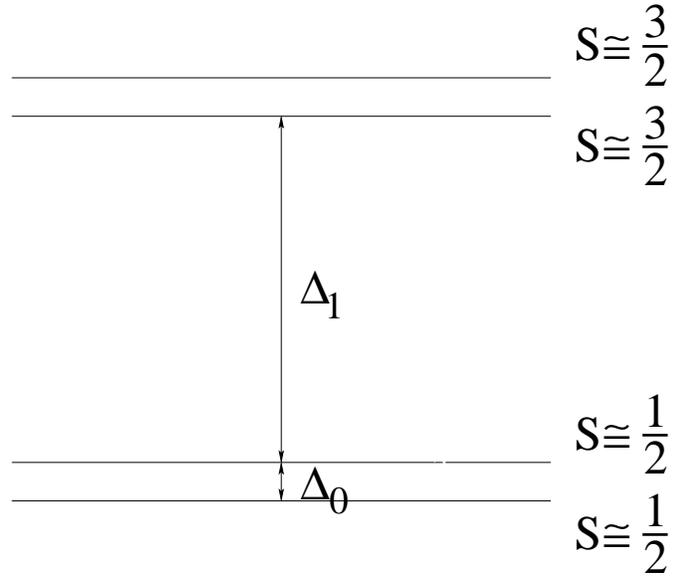,width=3.8in}
\end{center}
\vspace{20pt}
\caption{Low energy diagram for V$_{15}$.}
\label{fig:2}
\end{figure}

In equation (\ref{eqn:1}) the exchange constants given in \cite{Muller2} are
varied when $\lambda=1$ so that the gap from the doublets to the quartet agrees
with the experimental value of $\approx 3.7$ K. The following parameters are
kept:
\begin{equation}
J=1 \textrm{ , } J_{1} = J' = \frac{225}{800} J \textrm{ , } J_{2} = J'' =
\frac{350}{800} J
\end{equation}
The unit for $J$ is $800$ K and the energy gap to the excited state with these
values is equal to $\Delta_{1}=3.61$ K. Thus, the low energy diagram of
V$_{15}$ is as shown in figure \ref{fig:2}. $\Delta_{0} \sim 10^{-2}$ K is the
tunnel splitting between the two doublets. This splitting is not described by
the present Hamiltonian. A test for the success of perturbation theory is the
calculation of the expectation values of the square of the total spin
$\vec{S}^{2}$ and its projection on the $z$ axis, $S^{z}$, for the calculated
wavefunctions. These expectation values are half-integer for $S^{z}$ and
integer for $\vec{S}^{2}$ at $\lambda=1$ with a precision equal to the one used
to generate the perturbation expansions (double precision-15 digits). This
confirms the success of perturbation theory.

\subsection{Magnetic Anisotropy}
The anisotropic interaction between two spins $\vec{S}_{i}$ and $\vec{S}_{j}$
due to spin-orbit coupling is \cite{Dzyaloshinskii,Moriya}:
\begin{equation}
\vec{D}_{i,j} \cdot \vec{S}_{i} \times \vec{S}_{j} + \vec{S}_{i} \cdot
\mathit{\bf \Gamma_{i,j}} \cdot \vec{S}_{j}
\label{eqn:2}
\end{equation}
The first term is the DM interaction and it is first order in spin-orbit
coupling and antisymmetric. We refer to $\vec{D}_{i,j}$ as the DM vector. The
second term is second order in spin-orbit coupling and symmetric, with
$\mathit{\bf \Gamma_{i,j}}$ a second-rank tensor. Thus the DM vector provides
the lowest order anisotropy term.

As a minimal model, the DM interaction will be considered on the $J$ bonds
only, the strongest exchange constants in the system. The DM interaction is:
\begin{equation}
\vec{D}_{i,j} \cdot \vec{S}_{i} \times \vec{S}_{j} =
D_{i,j}^{x} ( S_{i}^{y} S_{j}^{z} - S_{i}^{z} S_{j}^{y} ) +
D_{i,j}^{y} ( S_{i}^{z} S_{j}^{x} - S_{i}^{x} S_{j}^{z} ) +
D_{i,j}^{z} ( S_{i}^{x} S_{j}^{y} - S_{i}^{y} S_{j}^{x} )
\end{equation}
Symmetry operations of the V$_{15}$ molecule, rotations of $\frac{2 \pi}{3}$
and $\frac{4 \pi}{3}$ around the axis that passes through the centers of the
hexagons ($z$), have to leave the form of the Hamiltonian invariant. This
constraints the $\vec{D}_{i,j}$ to be related to one another. If
\begin{equation}
\vec{D}_{1,2} = D_{1,2}^{x} \textrm{ } \hat{x} +
D_{1,2}^{y} \textrm{ } \hat{y} + D_{1,2}^{z} \textrm{ } \hat{z}
\end{equation}
then these three parameters determine the DM interaction in the other two $J$
bonds of the bottom hexagon as:
\begin{eqnarray}
\vec{D}_{3,4} = \frac{1}{2} \textrm{ } ( - D_{1,2}^{x} + \sqrt{3} \textrm{ } D_{1,2}^{y} ) \textrm{ } \hat{x} - \frac{1}{2} ( \sqrt{3} \textrm{ } D_{1,2}^{x} + D_{1,2}^{y} ) \textrm{ } \hat{y} + D_{1,2}^{z} \textrm{ } \hat{z} \nonumber\\
\vec{D}_{5,6} = - \frac{1}{2} \textrm{ } ( D_{1,2}^{x} + \sqrt{3} \textrm{ } D_{1,2}^{y} ) \textrm{ } \hat{x} + \frac{1}{2} ( \sqrt{3} \textrm{ } D_{1,2}^{x} - D_{1,2}^{y} ) \textrm{ } \hat{y} + D_{1,2}^{z} \textrm{ } \hat{z}
\end{eqnarray}
The sites $1, \cdots, 6$ are those in the bottom hexagon. Due to absence of
reflection symmetry with respect to the triangle plane the DM terms in the
upper hexagon will differ from the ones in the lower one. However, to minimize
the number of free parameters in the model, $\vec{D}_{10,11}=\vec{D}_{1,2}$.
Then from symmetry $\vec{D}_{12,13}=\vec{D}_{5,6}$ and
$\vec{D}_{14,15}=\vec{D}_{3,4}$.

In the same manner symmetry determines the elements of
$\mathit{\bf \Gamma_{i,j}}$. Considering the symmetric part of the spin-orbit
coupling again only on the strong bonds ($J$), and if
\begin{equation}
\mathit{\bf \Gamma_{1,2}} = \left( \begin{array}{ccc}
\Gamma_{1,2}^{xx} & \Gamma_{1,2}^{xy} & \Gamma_{1,2}^{xz}\\
\Gamma_{1,2}^{xy} & \Gamma_{1,2}^{yy} & \Gamma_{1,2}^{yz}\\
\Gamma_{1,2}^{xz} & \Gamma_{1,2}^{yz} & \Gamma_{1,2}^{zz}\\
\end{array} \right)
\end{equation}
then symmetry determines $\mathit{\bf \Gamma_{i,j}}$ by:
\begin{equation}
\mathit{\bf \Gamma_{i,j}} = U \textrm{ } \mathit{\bf \Gamma_{1,2}} \textrm{ }
U^{T}
\end{equation}
where $U$ is a rotation by $\phi$ around the $z$ axis:
\begin{equation}
U = \left( \begin{array}{ccc}
cos\phi & sin\phi & 0\\
-sin\phi & cos\phi & 0\\
0 & 0 & 1\\
\end{array} \right)
\end{equation}
$\mathit{\bf \Gamma_{3,4}}$ and $\mathit{\bf \Gamma_{5,6}}$ are determined in
this manner and, as in the case of the DM terms, to minimize the number of free
parameters in the model
$\mathit{\bf \Gamma_{10,11}}=\mathit{\bf \Gamma_{1,2}}$. Then from symmetry
$\mathit{\bf \Gamma_{12,13}}=\mathit{\bf \Gamma_{5,6}}$ and
$\mathit{\bf \Gamma_{14,15}}=\mathit{\bf \Gamma_{3,4}}$.



The DM term added to the $H_{1}$ term in equation (\ref{eqn:1}) is of the form:
\begin{eqnarray} 
H_{DM} = & \vec{D}_{1,2} \cdot \vec{S}_{1} \times \vec{S}_{2} +
\vec{D}_{3,4} \cdot \vec{S}_{3} \times \vec{S}_{4} +
\vec{D}_{5,6} \cdot \vec{S}_{5} \times \vec{S}_{6} + \nonumber \\
& \vec{D}_{10,11} \cdot \vec{S}_{10} \times \vec{S}_{11} +
\vec{D}_{12,13} \cdot \vec{S}_{12} \times \vec{S}_{13} +
\vec{D}_{14,15} \cdot \vec{S}_{14} \times \vec{S}_{15}
\end{eqnarray}
This term is considered as part of the perturbation, H$_1$, and is scaled with
the perturbation strength $\lambda$. From now on $D^{x} \equiv D_{1,2}^{x}$,
$D^{y} \equiv D_{1,2}^{y}$ and $D^{z} \equiv D_{1,2}^{z}$. This minimum model
is investigated for magnetic anisotropy and the parameters, $D^{x}$, $D^{y}$
and $D^{z}$, are determined by comparing with experiment.

We have found that the symmetric $\mathit{\bf \Gamma_{i,j}}$ terms, which are
second order in spin-orbit coupling, have little effect in the magnetization
curves. Once we have determined the parameters of the minimal model discussed
above, we will return to this point. In the case where the moments lie on an
open linear array it is possible to introduce local axes so that the DM term,
$\vec D_{ij} \cdot \vec S_i \times \vec S_j$, is absorbed into an effective
symmetric term \cite{Bonesteel}. However, the connectivity of the bonds between
the spins on the V sites ensures that this term can not be transformed away in
this manner.

\section{Results}

\label{sec:3}
\begin{figure}
\begin{center}
\epsfig{file=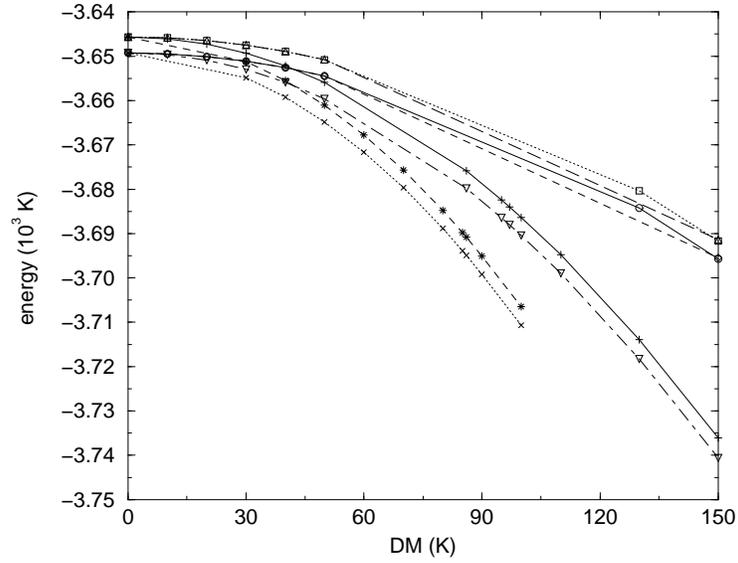,width=3.8in}
\end{center}
\vspace{20pt}
\caption{Average energy for the lowest and the highest multiplet in the
         $D^{x}$, $D^{y}$, $D^{z}$ parameter space: \newline (I)
         $D^{x}$=$D^{y}$=$0$, $D^{z}$, $\circ$: $\displaystyle S=\frac{1}{2}$,
         $\Box$: $\displaystyle S=\frac{3}{2}$ (II) $D^{x}$,
         $D^{y}$=$D^{z}$=$0$, $\diamond$: $\displaystyle S=\frac{1}{2}$,
         $\triangle$: $\displaystyle S=\frac{3}{2}$ \newline (III)
         $D^{x}$=$D^{y}$, $D^{z}$=$0$, $\bigtriangledown$:
         $\displaystyle S=\frac{1}{2}$, +: $\displaystyle S=\frac{3}{2}$ (IV)
         $D^{x}$=$D^{y}$=$D^{z}$ $\times$: $\displaystyle S=\frac{1}{2}$,
         $\ast$: $\displaystyle S=\frac{3}{2}$.}
\label{fig:3}
\end{figure}

\begin{figure}
\begin{center}
\epsfig{file=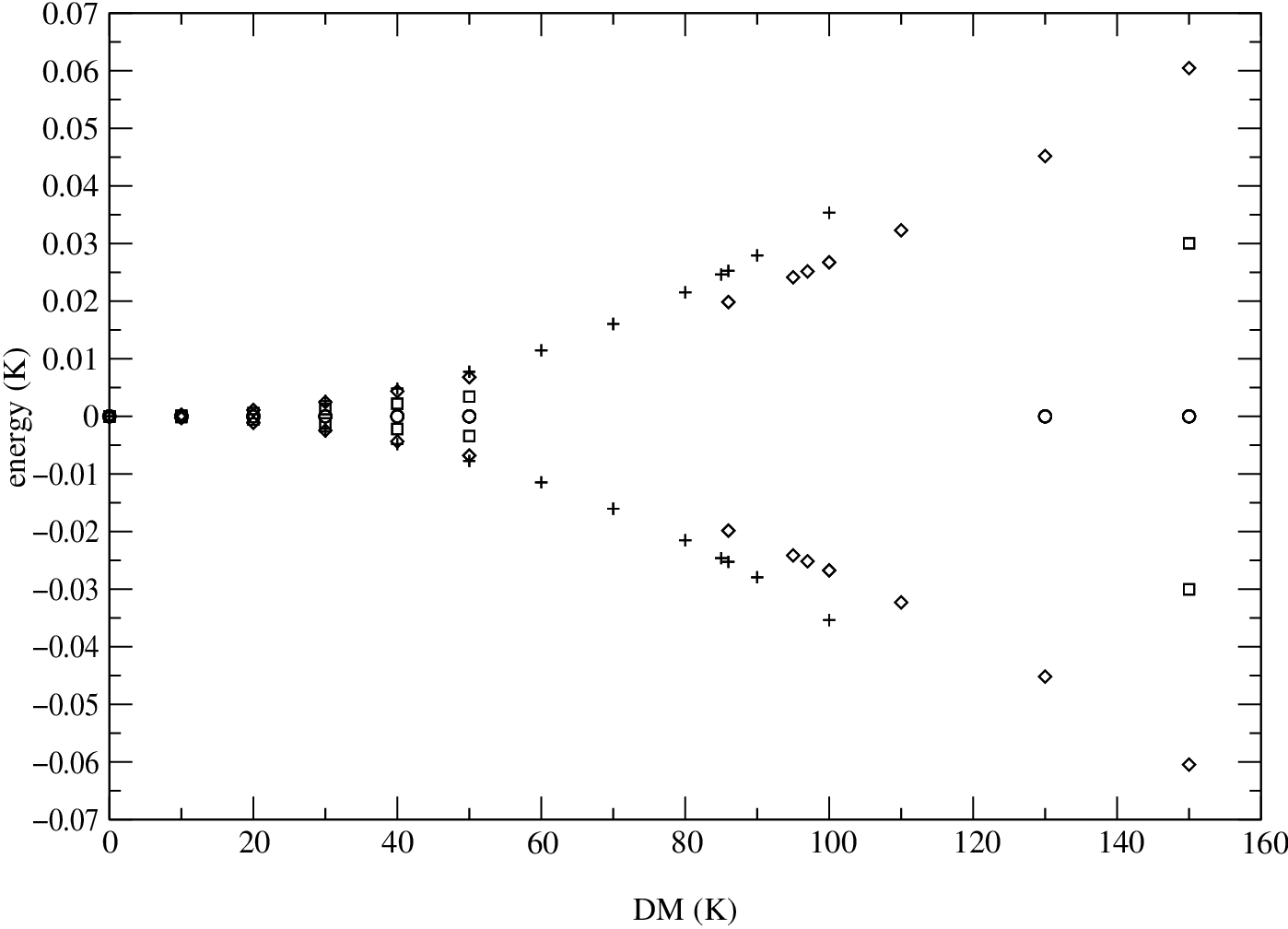,width=3.8in}
\end{center}
\vspace{20pt}
\caption{Deviations from the average energy for the two doublets in the
         $D^{x}$, $D^{y}$, $D^{z}$ parameter space: \newline (I) $\circ$:
         $D^{x} = D^{y} = 0$, $D^{z}$ (K) (II) $\Box$: $D^{x}$ (K),
         $D^{y} = D^{z} = 0$ \newline (III) $\diamond$: $D^{x} = D^{y}$ (K),
         $D^{z} = 0$ (IV) +: $D^{x} = D^{y} = D^{z}$ (K).}
\label{fig:4}
\end{figure}

\begin{figure}
\begin{center}
\epsfig{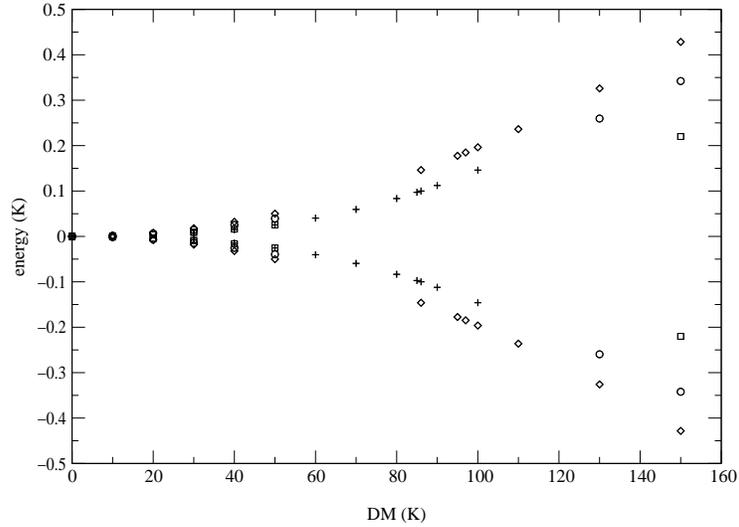}
\end{center}
\vspace{20pt}
\caption[Deviations from the average energy for the quartet in the $D^{x}$,
         $D^{y}$, $D^{z}$ parameter space.]{Deviations from the average energy
         for the quartet in the $D^{x}$, $D^{y}$, $D^{z}$ parameter space:
         \newline (I) $\circ$: $D^{x} = D^{y} = 0$, $D^{z}$ (K) (II) $\Box$:
         $D^{x}$ (K), $D^{y} = D^{z} = 0$ \newline (III) $\diamond$:
         $D^{x} = D^{y}$ (K), $D^{z} = 0$ (IV) +: $D^{x} = D^{y} = D^{z}$ (K).}
\label{fig:5}
\end{figure}

The effect of the DM terms in the energy spectra is investigated by varying the
strength of $D^{x}$, $D^{y}$ and $D^{z}$. The average energies of the two
doublets and the quartet are plotted for different values of the $\vec{D}$
vector in figure \ref{fig:3}. The changes in the energy become more pronounced
as $|\vec{D}|$ increases. However, one sees that even DM terms $\sim 100$ K
lead to energy shifts $\sim 20$ K. The deviations from the average energies of
the two doublets are plotted in figure \ref{fig:4}, while for the quartet in
figure \ref{fig:5}. It can be seen that even when $|\vec{D}| \sim 100$ K the
sizes of the energy splittings in the two doublets and the quartet are modest.
This suggests that the energy level differences estimated from the experiment
are comparatively insensitive to $\vec D$ so that it can not alone determine
$\vec{D}$.

These spectra arise from the form of the effective Hamiltonian for the triange
spins, $H^{eff}$, calculated with perturbation theory. The effective
Hamiltonian for the triangle spins is of the form
$H^{eff}=H^{diag}+H^{off-diag}$, where $H^{diag}$ is a constant diagonal matrix
in the basis of total $S^{z}$ states which is three orders of magnitude larger
than the terms in $H^{off-diag}$ and depends on the magnitude of the anisotropy
term, as can be seen in figure \ref{fig:3}. All the total $S^{z}$ sectors of
the triangle spins are coupled in $H^{off-diag}$ even when the anisotropy term
is $\vec D^{z} \cdot \sum_{ij}\vec S_{i} \times \vec S_{j}$ on the $800$ K
hexagonal bonds, case (I) in figure \ref{fig:4} and figure \ref{fig:5}.

By contrast, when one considers a three-spin model with a Heisenberg coupling,
$J_{0} \sum_{ij} \vec{S}_{i} \cdot \vec{S}_{j}$, and an anisotropy term,
$\vec D^{z} \cdot \sum_{ij}\vec S_{i} \times \vec S_{j}$, between the triangle
spins, the quartet remains degenerate while the doublets are split into
time-reversed pairs separated by $\sqrt{3} D^{z}$. Unlike the result of this
three-spin model, the energies of all quartet levels are changed by the
anisotropy term, as can be seen from figures  \ref{fig:4} and \ref{fig:5}, and
the splitting of the higher energy multiplet is larger than the low energy one.
This points to the difficulty in understanding the effect of the anisotropy
terms on $H^{eff}$ in terms of a simple model for the triangle spins. More
generally it supports the conclusions of De Raedt et al. \cite{DeRaedt00} on
the dangers of Hilbert space truncation.

\subsection{Magnetization}

The magnetization of V$_{15}$ was calculated for various magnetic fields and
temperatures as a function of the DM terms. In the absence of DM terms and at
zero temperature, $S^{z}$ is a good quantum number and the magnetization stays
constant except at critical fields where there are level crossings. In the
first level crossing the ground state selects the state of the zero field
ground state doublet with the higher expectation value of the spin, while in
the second it switches from this doublet to one of the quartet states. When the
temperature differs from zero, the jumps are smeared out and at high enough
temperatures there are no sudden changes in the magnetization.

\begin{figure}
\begin{center}
\epsfig{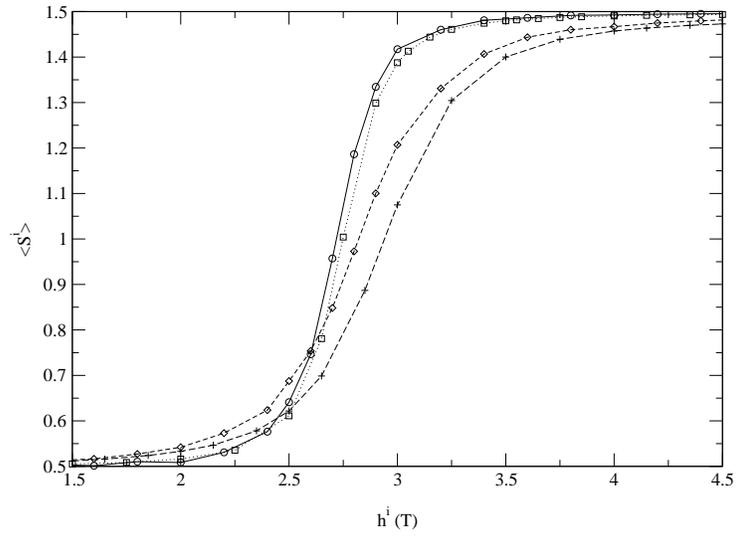}
\end{center}
\vspace{20pt}
\caption[$<S^{i}>$ vs. $h^{i}$, $D^{y} = D^{z} = 0$, T = 0.001 K.]{$<S^{i}>$
         vs. $h^{i}$, $D^{y} = D^{z} = 0$, T = 0.001 K: $\circ$: $i=x$,
         $D^{x} = 50$ K, $\Box$: $i=z$, $D^{x} = 50$ K, $\diamond$: $i=x$,
         $D^{x} = 100$ K, +: $i=z$, $D^{x} = 100$ K.}
\label{fig:6}
\end{figure}

When the DM terms are non-zero at zero temperature $\vec{S}^{2}$ does not
commute with the Hamiltonian, and the same is true for $S^{z}$ if either
$D^{x}$ or $D^{y}$ is non-zero. The magnetization along the field was
calculated for $D^{x} = 50$ K and $D^{x} = 100$ K at a very low temperature,
$T=0.001$ K, in the case of a field along the $x$ axis and a field along the
$z$ axis and is plotted in figure \ref{fig:6}. The gyromagnetic ratio used is
$g=2.0023$ \cite{Kaku}. It observed that the magnetization depends on the
degree of anisotropy, $D^{x}$. In addition, it is seen that the jump is more
smeared out as the magnitude of $D^{x}$ increases from $0$ to $100$ K. Magnetic
anisotropy is demonstrated by the dependence of the magnetization along the
direction of the field on the direction of the field itself.

\begin{figure}
\begin{center}
\epsfig{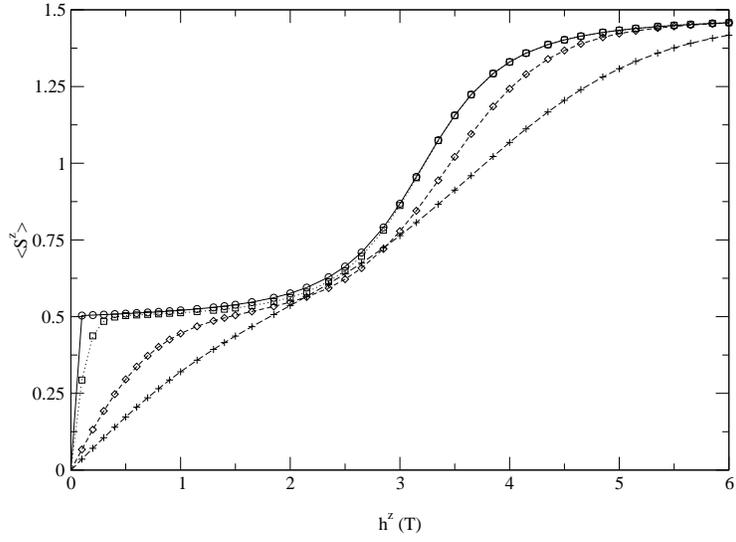}
\end{center}
\vspace{20pt}
\caption[$<S^{z}>$ vs. $h^{z}$, $D^{x} = D^{y} = D^{z} = 100$ K.]{$<S^{z}>$ vs.
         $h^{z}$, $D^{x} = D^{y} = D^{z} = 100$ K, $\circ$: T = 0.001 K,
         $\Box$: T = 0.1 K, $\diamond$: T = 0.5 K, +: T = 1 K.}
\label{fig:7}
\end{figure}

The broadening of the spin transition can also be observed in figure
\ref{fig:7}. There the magnetization $<S^{z}>$ is plotted as a function of the
magnetic field for four different temperatures and a DM term of magnitude
$\sqrt{3}$ $100$ K. The magnetization curve is almost identical for $T=0.001$ K
and $T=0.1$ K. This shows that the broadening of the transition is not a
temperature effect, but it is associated with the DM terms. For high enough
temperatures the susceptibility is almost constant with the field.

\begin{figure}
\begin{center}
\epsfig{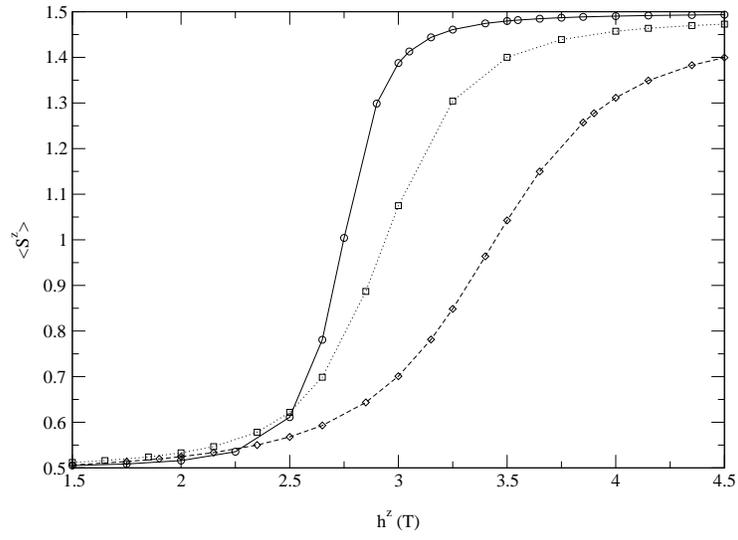}
\end{center}
\vspace{20pt}
\caption[$<S^{z}>$ vs. $h^{z}$, T = 0.001 K, $D^{y} = D^{z} = 0$ K.]{$<S^{z}>$
         vs. $h^{z}$, T = 0.001 K, $D^{y} = D^{z} = 0$ K, $\circ$: $D^{x}$ = 50
         K, $\Box$: $D^{x}$ = 100 K, $\diamond$: $D^{x}$ = $\sqrt{3}$ 100 K.}
\label{fig:8}
\end{figure}

\begin{figure}
\begin{center}
\epsfig{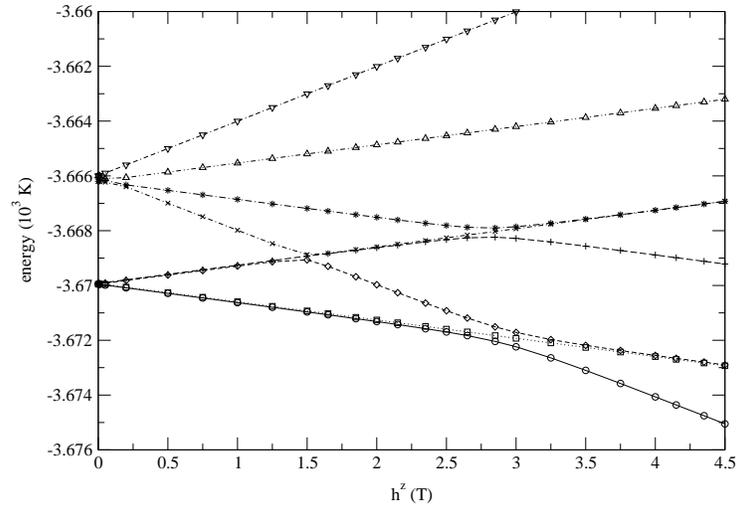}
\end{center}
\vspace{20pt}
\caption{Energies of the eight-state manifold vs. $h^{z}$, $D^{x}$=$100$ K,
         $D^{y}$=$D^{z}$=$0$ K.}
\label{fig:9}
\end{figure}

In figure \ref{fig:8} the calculated magnetization is plotted for three
different values of $D^{x}$ with $D^{y} = D^{z} = 0$. It is seen that the width
of the transition gets bigger with the increase of the DM terms. The origin of
the width can be seen in the energy spectrum plotted in figure \ref{fig:9}. The
presence of matrix elements between the states in the vicinity of the
transition field ($\sim 3$ T) leads to level anticrossings. Pairs of degenerate
states from the lower multiplet in zero field split up with one hybridizing
with a member of the quartet multiple with $S^{z} \approx \frac{3}{2}$, while
the other has no matrix element with this state.

\begin{figure}
\begin{center}
\epsfig{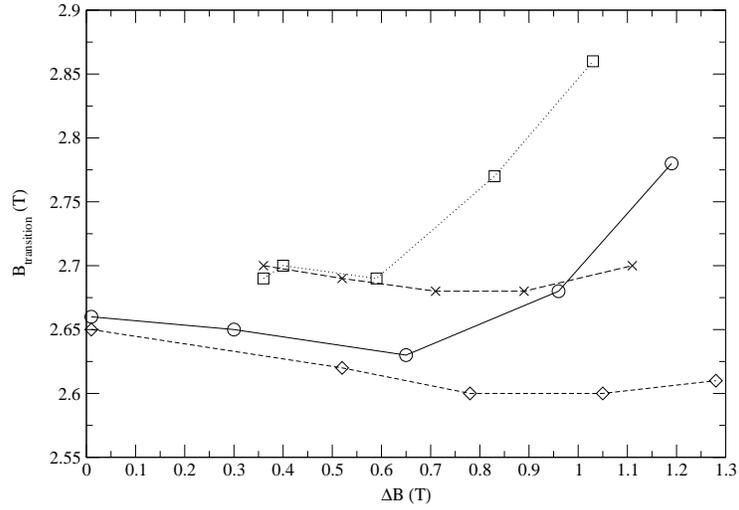}
\end{center}
\vspace{20pt}
\caption{Midpoint $B_{transition}$ versus width $\Delta B$ of the magnetization
         transition. \newline (I) $D^{y}$=$D^{z}$=$0$, left to right:
         $D^{x} = 0$, $25$, $50$, $80$, $100$ K,  $\circ$: $T = 10^{-3}$ K,
         $\Box$: $T = 0.1$ K \newline (II) $D^{z}$=$100$ K, left to right:
         $D^{x}$=$D^{y}$=$0$, $20$, $30$, $40$, $50$ K, $\diamond$:
         $T = 10^{-3}$ K, $\times$: $T = 0.1$ K.}
\label{fig:10}
\end{figure}

The width of the magnetization transition is an indication of the effect of the
DM terms, as can be seen from figures \ref{fig:7} and \ref{fig:8}. The
transition is characterized by its midpoint $B_{transition}$ and
width-difference $\Delta B$ in magnetic field values where it has changed by 10
$\%$ from its low and high field average spin values on either side of the
transition. The results for various DM terms are shown in table \ref{table:1},
where it can be seen that $B_{transition}$ depends sensitively on the form of
the DM terms. This dependence on the DM terms is illustrated in figure
\ref{fig:10} where we plot $B_{transition}$ versus $\Delta B$ for two different
forms of the DM vector, one in which $D^{x} \neq 0$ K, $D^{y}=D^{z}=0$ and the
other in which $D^{z} \neq 0$ K, $D^{x}=D^{y}=0$. The values of
$B_{transition}$ and the corresponding values of $\Delta B$ shown are for
different magnitudes of $D^{x}$ and $D^{z}$ at two different temperatures.
Although the magnetization curves for these two temperatures appear to lie on
top of one another in figure \ref{fig:7}, there is a difference which should be
taken into account in quantitative comparison with data. The effect of
increasing temperature is seen to push $B_{transition}$ to higher fields, while
$\Delta B$ is seen to be non-monotonic. For small magnitudes of the DM term
$\Delta B$ is larger at $T=0.1$ K than at $T=10^{-3}$ K but this is reversed
for a sufficiently large magnitude depending on the two forms considered.

These temperature dependences can be understood as follows. $B_{transition}$
moves to higher applied fields with increasing temperature because the higher
energy low magnetization state contributes significantly to the thermodynamic
average for larger energy differences between the ground and excited state,
that is for larger applied fields beyond the field at which the two states are
degenerate. The temperature dependence of $\Delta B$ is more complicated. Both
the temperature and the $\vec D_{ij}$ vectors contribute to $\Delta B$. Because
the temperatures considered here are so low compared to the variation in the
magnitudes of the $\vec D_{ij}$ vectors, the difference in the smearing effects
of the different temperatures, $\simeq 0.1$ K, can be neglected compared to
that due to the hybridization of the levels near $B_{transition}$. The increase
in $\Delta B$ for $T=10^{-3}$ K compared to $T=0.1$ K arises because the
transition takes place at lower values of the applied field at $T=10^{-3}$ K,
$\sim 2.60$ T, instead of $\sim 2.68$ T at $T=0.1$ K, and because the effect of
the $\vec D_{ij}$ vectors is larger at lower values of the applied field once
the energy levels are close. The matrix elements associated with the
$\vec D_{ij}$ vectors are larger compared to the Zeeman energies in the
effective Hamiltonian. As a result increasing the magnitude of $\vec D_{ij}$
vectors leads to a larger effect on $\Delta B$ at lower temperatures. It was
also observed that the results do not depend on the sign of the DM terms.

The above results were calculated with a gyromagnetic ratio $g=2.0023$. A
similar analysis for experimentally determined temperature dependence of the
magnetization at $T=0.1$ K gives $\Delta B = 0.77 \pm 0.10$ T and
$B_{transition} = 2.84 \pm 0.05$ T \cite{Chiorescu}. In order to compare with
this data, an average of the values found from EPR spectra for the gyromagnetic
ratio is used \cite{Muller3}. They are $g_{a}=g_{b}=1.95$ and $g_{c}=1.98$,
where $a$, $b$ and $c$ are the crystallographic axes of the molecule. Using
$g=1.97$ and $D^{x}=70$ K, $D^{y}=D^{z}=0$, it is found that $\Delta B = 0.75$
T and $B_{transition} = 2.81$ T, which is within the uncertainty of the
experimental value. Therefore, the inclusion of DM terms accounts for the
rounding of the magnetization at very low temperatures and provides a
description of the magnetic anisotropy of V$_{15}$. In addition, the gap
between the doublets and quartet, $\sim 3.7$ K, is consistent with the
estimates given by Chiorescu et al. \cite{Chiorescu} and Chaboussant et al.
\cite{Chab02} for the effective coupling of the spins in the triangle, induced
by interactions between these and the spins in the hexagons. Both groups found
$J_{0} \simeq 2.4$ K which gives a gap $\frac{3}{2} J_{0} \simeq 3.7$ K. We
also found that the tunnel splitting between the two doublets is
$\Delta_{0} = 13$ mK, in agreement with the $\sim 10^{-2}$ K estimate
\cite{Carter}. This is an order of magnitude smaller than the estimates given
by Chiorescu et al. \cite{Chiorescu2} ($\sim 0.05$ K) or Chaboussant et al.
($\sim 0.1$ K). We note that the calculated splittings of the quartets, shown
in figure \ref{fig:5}, are an order of magnitude larger than those of the
doublets, shown in figure \ref{fig:4}, which may account for the discrepancy
between experimental estimates and our results.

In arriving at the form and magnitude of the $\vec D_{ij}$ we have relied on a 
simple model in which the $\vec D_{ij}$ are non-zero only on the strongest
bonds and it is not obvious how model dependent our results are in the absence
of more detailed data. The original estimates by Moriya \cite{Moriya} relied on
a superexchange model where magnetic order is driven by strong short-range
correlations due to strongly suppressed hopping between sites. Estimates of the
$\vec D_{ij}$'s then come from considering the mixing of orbitals by angular
momentum matrix elements. In V$_{15}$ interactions between ions on hexagon
sites are mediated by pairs of oxygen atoms. Depending on whether one of these
oxygen atoms is linked to an arsenic atom or another vanadium atom, the
Heisenberg coupling is either $800$ K or $150$ K. In these circumstances it is
not clear how reliable estimates are based on the simple tight-binding
approach. The magnitude of the $\vec D_{ij}$ vectors found here is comparable
to the Heisenberg couplings. This is larger than the estimate based on Moriya's
original approach which gives
${{|\vec D_{ij}|}\over{J}}\sim {{\Delta g}\over{g}}$, where $\Delta g$ is the
shift in the gyromagnetic ratio from that of a free electron. However,
Katsnelson et al. \cite{Katsnelson99} also found strong Dzyaloshinskii-Moriya
interactions with values comparable to the isotropic constants in their
analysis, which was based on an effective eight spin model of neutron
scattering and EPR data on Mn$_{12}$.

\begin{figure}
\begin{center}
\epsfig{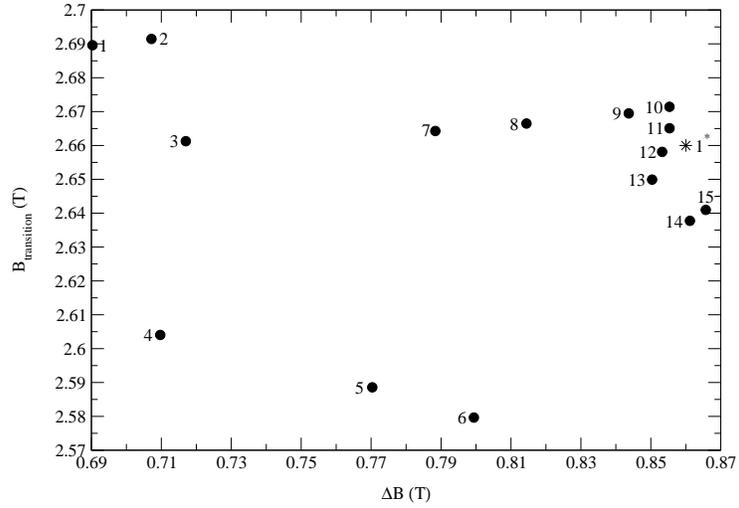}
\end{center}
\caption{Effect of $\mathit{\bf \Gamma_{i,j}}$ on $B_{transition}$ and
$\Delta B$. Each point corresponds to a different $\mathit{\bf \Gamma_{i,j}}$
according to table \ref{table:2}. $\mathit{\bf \Gamma_{i,j}}=\mathit{\bf 0}$
for point $1^{*}$.}
\label{fig:13}
\end{figure}

The determination of $\vec D_{ij}$ is carried out in the absence of the
symmetric anisotropy term, $\mathit{\bf \Gamma_{i,j}}$. We investigated the
effect of including $\mathit{\bf \Gamma_{i,j}}$'s with a range of magnitudes
for its elements with different $\vec D_{ij}$'s. Since the $\vec D_{ij}$'s are
first order in the spin-orbit interaction while the
$\mathit{\bf \Gamma_{i,j}}$'s are second order, we have assumed the magnitudes
of the symmetric anisotropy terms to be smaller than the antisymmetric ones.
For illustration we show in fig. \ref{fig:13} the dependence of
$B_{transition}$ and the corresponding values of $\Delta B$ for different forms
of $\mathit{\bf \Gamma_{i,j}}$ given in table \ref{table:2}. In this case
$D^{x}=70$ K, $D^{y}=D^{z}=0$ with $g=2.0023$. One sees that the values of
$B_{transition}$ and $\Delta B$ are not significantly changed from the values
given by the DM terms alone. It was observed here that the results depend on
the sign of the ${\bf \Gamma_{i,j}}$ terms.


\subsection{Magnetic Relaxation}

Apart from equilibrium magnetization measurements, the magnetization of
V$_{15}$ in a varying external field has been measured. The data shows a
so-called ``butterfly'' hysteresis, where the magnetization initially increases
with the field, then reaches a plateau, and eventually approaches saturation
\cite{Chiorescu,Chiorescu2}. This has been attributed to non-equilibrium
distribution of phonons, and the spin-phonon mechanism for magnetic relaxation
is called ``hole-burning''. Here the role of DM terms in magnetic relaxation is
investigated.

In order to calculate the magnetization in a swept field as in the experiments
it is necessary to calculate the density matrix for the eight low energy
states. However, this is a prohibitively expensive numerical calculation if one
is to reach applied fields of $7$ T with the sweeping rate of $0.14$ T/s. This
is because the characteristic time,
$\tau_{0}= \frac{\hbar}{k_{B} K} = 7.64 \times 10^{-12}$ s, associated with
anisotropy terms is so short compared to the $\sim 50$ s required to reach $7$
T, that the number of time steps needed is very large. The presence of rapidly
varying phases in the elements of the density matrix requires a very fine time
step $\ll \tau_{0}$. The computational effort to get accurate results is
enormous. Therefore, rather than consider a time dependent magnetic field which
changes at a constant rate, we consider the case in which a $10$ T field along
the $z$ direction is suddenly switched on, off, or reversed.
In the absence of magnetic anisotropy there is no change in $<S^{z}>$ with
time. However, this is not the case in the presence of DM terms. As can be seen
in figure \ref{fig:9}, there is hybridization between the zero field states as
a function of applied field, and matrix elements responsible for this lead to a
time dependent $<S^{z}>$ when the applied field is suddenly changed.

We illustrate this point by calculating the magnetization for the case where
the magnetic field has an initial strength of $10$ T, and $D^{x} = 80$ K,
$D^{y} = 30$ K, $D^{z} = 0$ and $g=2.0023$. The ground state magnetization is
$<S^{z}>=1.49$. Then the magnetic field is suddenly switched off, and the
evolution of the magnetization is calculated in the basis of the eight lowest
states of the same problem in zero magnetic field. Any excited state higher in
energy is separated by a gap of order $J$ from the lowest energy manifold, and
the probability of the state of the system belonging to this manifold is
$0.9997$, so this approximation is well justified.

\begin{figure}
\begin{center}
\epsfig{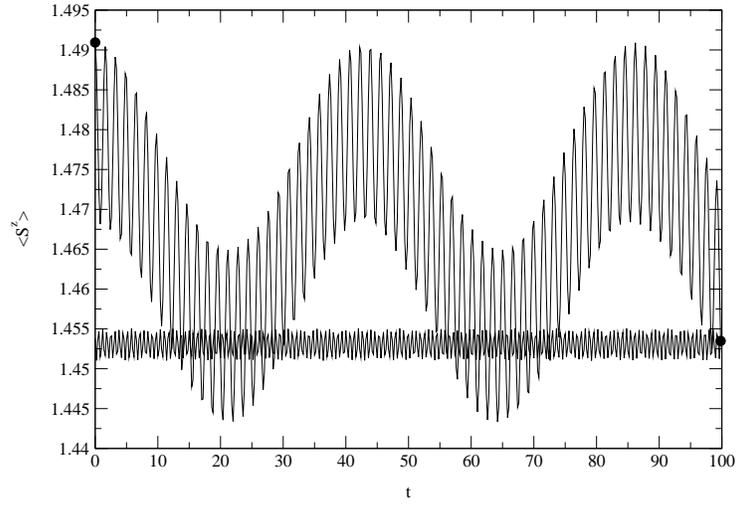}
\end{center}
\vspace{20pt}
\caption{$<S^{z}>$ vs. time, time in units of
         $\tau_{0}=\frac{\hbar}{k_{B} K} = 7.64 \times 10^{-12}$ s, $T=0$,
         $D^{x} = 80$ K, $D^{y} = 30$ K, $D^{z} = 0$ K: ``upper'' solid line:
         after the field is switched off,
         ``lower'' solid line: field switched back on to $10$ T.
         The black circles indicate $<S^{z}>$ at the moments where the field is
         switched off and back on.}
\label{fig:14}
\end{figure}

The magnetization varies harmonically with time, as seen in figure
\ref{fig:14}, and the frequencies of its variation are determined by the energy
differences of the system at zero magnetic field. After a time equal to
$100 \tau_0 \simeq 10^{-10}$ s, the field is turned back on to $10$ T and
$<S^{z}>$ now varies with different frequencies and amplitude. The presence of
more than one frequency in the time dependence points to the inadequacy of a
two-level model and suggests that there are significant corrections to any
Landau-Zener treatment of the relaxation. The average value of the
magnetization is different in the two cases, and in particular the average
value after the field is turned on again depends on the wavefunction at the
exact moment the field is switched. The origin of this effect is the magnetic
anisotropy of the DM terms, a source of magnetic decoherence for the molecule.
The calculation can also be done starting from the zero field ground states in
a non-zero magnetic field, and after $\sim 10^{-10}$ s switching off the field
suddenly. The time dependence is similar to the previous case, and once again
the average value of $<S^{z}>$ after the field has been switched off depends on
the state of the system at the moment it was switched off. The DM terms
therefore provide a source of spin decoherence with a decoherence time
$\sim 10^{-12}$ s.

In both cases the calculation was done assuming instantaneously switched on and
off fields and demonstrates magnetic decoherence and hysteresis in the presence
of DM terms. If a time varying field is introduced in the calculation, the
presence of rapidly varying phases requires a very fine time step for the
calculation of the time evolution of the magnetization. This time step has to
be a fraction of $10^{-12}$ s, and a calculation of the evolution of the
magnetization for realistic times would require an enormous amount of
computational effort. However, the small time scale calculation demonstrates
the intrinsic magnetic decoherence in the presence of magnetic anisotropy.
Hysteresis is seen in experiment and attributed to the spin-phonon interaction
\cite{Chiorescu}. The present mechanism is independent of temperature and can
be expected to provide the dominant relaxation process at very low
temperatures. 

\section{Conclusion}

\label{sec:4}
Degenerate perturbation theory was used to study magnetic anisotropy in the
molecule V$_{15}$. 
%
The DM term introduced in the model quantitatively accounts for the rounding of
transitions in the magnetization at very low temperatures, their width at
midpoint, through the avoided level crossing, as well as observed zero field
splitting of the lowest four states.

We have shown that the DM interaction is also responsible for spin decoherence
and magnetic hysteresis. This is present even at very low temperatures where
thermal mechanisms are absent. This source of decoherence may pose a problem
for the application of magnetic molecules in quantum computing, since it is
necessary to control the time evolution of states with external magnetic fields
and electromagnetic pulses \cite{Leuenberger}. In order for a molecule with a
DM interaction to be suitable for quantum computing it would be necessary to
carry out operations on a time scale much shorter than
$\frac{\hbar}{\Delta E}$, where $\Delta E$ is comparable to the magnitude of
the DM interaction. In the present example this is $10^{-12}$ s.

The use of perturbation theory to calculate the low energy and temperature
properties of V$_{15}$ is an alternative to exact diagonalization which has
been discussed by Honecker et al. \cite{Honecker} for Fe$_{6}$ and Fe$_{8}$.
Perturbation theory works particularly well in the case of V$_{15}$ because of
the strength of the hexagonal bonds. However, we have also applied perturbation
theory to a system of $20$ $S=\frac{1}{2}$ spins by calculating the effect of
quantum fluctuations around the classical ground state\cite{NPK}. The dimension
of the Hilbert space in that case is almost exactly the same as Fe$_8$,
$\sim 10^6$. It is possible that this approach will also succeed in systems
with larger, more classical moments and larger Hilbert spaces. This would allow
a more microscopic approach to the treatment of Mn$_{12}$, which may resolve
some of the questions raised by Zhong et al. \cite{Sarachik}.

\section{Acknowledgements}

This work was carried out at the Center for Computational Research at SUNY
Buffalo. N.P.K. is supported by a Marie Curie Fellowship of the European
Community program Development Host Fellowship under contract number
HPMD-CT-2000-00048.

$^{*}$ Current address: Department of Physics, Buffalo State College, 1300
Elmwood Avenue, Buffalo, NY 14222.

%
%

%
%

\newpage

\begin{table}
\begin{center}
\caption{Magnetization transition. Index $A$ refers to $T = 0.001$ K, while $B$
to $T = 0.1$ K.}
\begin{tabular}{c|c|c|c|c|c|c|c|c}
$D^{x}$ (K) & $D^{y}$ (K) & $D^{z}$ (K) & $\Delta_{0}$ (K) & $\Delta_{1}$ (K) &
$\Delta B_{A}$ (T) & $B_{tA}$ (T) & $\Delta B_{B}$ (T) & $B_{tB}$ (T)\\
\hline
  0 &  0 &   0 & 0     & 3.611 & 0.08 & 2.66 & 0.36 & 2.69 \\
\hline
 25 &  0 &   0 & 0.002 & 3.618 & 0.30 & 2.65 & 0.40 & 2.70 \\
\hline
 50 &  0 &   0 & 0.007 & 3.641 & 0.65 & 2.63 & 0.59 & 2.69 \\
\hline
 80 &  0 &   0 & 0.017 & 3.688 & 0.96 & 2.68 & 0.83 & 2.77 \\
\hline
100 &  0 &   0 & 0.027 & 3.730 & 1.19 & 2.78 & 1.03 & 2.86 \\
\hline
  0 &  0 & 100 & 0     & 3.618 & 0.07 & 2.65 & 0.36 & 2.70 \\
\hline
 20 & 20 & 100 & 0.003 & 3.648 & 0.52 & 2.62 & 0.52 & 2.69 \\
\hline
 30 & 30 & 100 & 0.007 & 3.685 & 0.78 & 2.60 & 0.71 & 2.68 \\
\hline
 40 & 40 & 100 & 0.013 & 3.736 & 1.05 & 2.60 & 0.89 & 2.68 \\
\hline
 50 & 50 & 100 & 0.020 & 3.801 & 1.28 & 2.61 & 1.11 & 2.70 \\
\end{tabular}
\label{table:1}
\end{center}
\end{table}

\begin{table}
\begin{center}
\caption{Magnetization transition with inclusion of $\mathit{\bf \Gamma_{i,j}}$
terms. $D^{x}=70$ K, $D^{y}=D^{z}=0$ K, $T=0.001$ K.}
\begin{tabular}{c|c|c|c|c|c|c|c|c}
point & $\Gamma_{1,2}^{xx}$ (K) & $\Gamma_{1,2}^{yy}$ (K) & $\Gamma_{1,2}^{zz}$
(K) & $\Gamma_{1,2}^{xy}$ (K) & $\Gamma_{1,2}^{xz}$ (K) & $\Gamma_{1,2}^{yz}$
(K) & $\Delta B$ (T) & $B_{t}$ (T)\\
\hline
 $1^{*}$ &  0 &  0 &  0 &  0 &  0 &  0 & 0.86 & 2.66 \\
\hline
       1 &  0 &  0 &  0 &  0 & 50 &  0 & 0.69 & 2.69 \\
\hline
       2 &  0 &  0 &  0 & 50 & 50 &  0 & 0.71 & 2.69 \\
\hline
       3 &  0 &  0 &  0 & 50 & 50 & 50 & 0.72 & 2.66 \\
\hline
       4 &  0 &  0 & 50 & 20 & 20 & 20 & 0.71 & 2.60 \\
\hline
       5 & 10 & 10 & 10 & 10 & 10 & 10 & 0.77 & 2.59 \\
\hline
       6 & 10 & 10 & 10 &  0 &  0 & 20 & 0.80 & 2.58 \\
\hline
       7 &  0 &  0 &  0 & 20 & 20 & 20 & 0.79 & 2.66 \\
\hline
       8 &  0 &  0 &  0 & 10 & 10 & 10 & 0.81 & 2.67 \\
\hline
       9 &  0 &  0 &  0 & 10 &  0 &  0 & 0.84 & 2.67 \\
\hline
      10 &  0 &  0 &  0 & 50 &  0 &  0 & 0.86 & 2.67 \\
\hline
      11 &  0 &  0 &  0 & 50 &  0 & 10 & 0.86 & 2.67 \\
\hline
      12 &  0 &  0 &  0 & 40 &  0 & 20 & 0.85 & 2.66 \\
\hline
      13 &  0 &  0 &  0 &  0 &  0 & 30 & 0.85 & 2.65 \\
\hline
      14 &  0 &  0 &  0 &  0 &  0 & 50 & 0.86 & 2.64 \\
\hline
      15 &  0 &  0 &  0 & 50 &  0 & 50 & 0.87 & 2.64 \\
\end{tabular}
\label{table:2}
\end{center}
\end{table}

%
%

\newpage

\vspace{10pt}

\vspace{10pt}

\vspace{10pt}

\vspace{10pt}



\vspace{10pt}

\end{document}